\documentclass[journal]{IEEEtran}
\usepackage{amsmath,amsfonts}
\usepackage{algorithm}
\usepackage{array}
\usepackage[caption=false,font=normalsize,labelfont=sf,textfont=sf]{subfig}
\usepackage{textcomp}
\usepackage{stfloats}
\usepackage{url}
\usepackage{verbatim}
\usepackage{graphicx}
\usepackage{tikz}
\usepackage{cite}
\usepackage[useregional]{datetime2}

\usepackage{graphicx}
\usepackage{booktabs}
\usepackage{array}
\usepackage{multirow}

\usepackage[table]{xcolor}
\usepackage{colortbl}

\definecolor{FFcolor}{RGB}{210,235,255}   
\definecolor{MMcolor}{RGB}{255,240,200}   
\definecolor{MFcolor}{RGB}{210,255,210}   

\newcommand{\spk}{\mathrm{spk}}
\newcommand{\famlab}{\mathrm{fam}}
\newcommand{\aux}{\mathrm{aux}}

\renewcommand{\vec}[1]{\boldsymbol{\mathrm{#1}}}

\usepackage{enumitem}
\usepackage{xcolor}
\usepackage{tikz}

\usepackage{algpseudocode}

\begin{document}

\title{Kinship Verification Using Voice} 

\author{Jagabandhu Mishra, and Tomi H. Kinnunen
\thanks{ 
J. Mishra (jagabandhu.mishra@uef.fi) and T.H. Kinnunen (tomi.kinnunen@uef.fi) are with the University of Eastern Finland, Joensuu, Finland. \\
This study has been partially supported by the Academy of Finland (Decision No. 349605, project "SPEECHFAKES").}

}


\markboth{Journal of  Class Files,~Vol.~14, No.~8, August~2023}%
{Shell \MakeLowercase{\textit{et al.}}: A Sample Article Using IEEEtran.cls for IEEE Journals}


\maketitle

\begin{abstract}
Kinship verification (KV) from voice, the task of determining whether two
speakers are biologically related, has received only little attention. Our work establishes a foundational basis for this emerging frontier, contributing to both performance evaluation and detection methodologies. First, leveraging the speech recordings of the large-scale audio-visual dataset, KAN-AV, we propose a revised evaluation protocol that controls for various confounders and adopts a family-disjoint train--test split to address open-set KV. Second, we analyze the close connection between speaker verification and KV, showing that genealogical similarity of speaker pairs plays opposite roles in the two tasks. Third, we tackle KV using three neural speaker embedding extractors (ECAPA-TDNN, WavLM-ECAPA, and ReDimNet) combined with various back-ends. In zero-shot KV including same-speaker target trials, ReDimNet achieves the lowest equal error rate (EER) of $20.8\%$; however, performance degrades to $39.7\%$ under strict kin trials, where same-speaker target trials are excluded. Our best trainable back-end, which applies asymmetric processing of the embedding pair to mitigate age-difference effects, obtains an EER of $32.0\%$ ($18.6\%$ with speaker target trials included). These results highlight the difficulty of KV while showing that speaker embeddings encode familial cues, offering a promising foundation for voice-based kinship analysis.
\end{abstract}

\section{\label{sec:1} Introduction}

\IEEEPARstart{K}{inship} refers to biological relationships among individuals connected by blood, such as parent-child and brother-sister~\cite{schneider1972kinship}. Owing to its importance for understanding family structures and inheritance patterns, as well as its practical relevance in forensics, \emph{kinship verification} (KV) has been studied across multiple disciplines, including genetics~\cite{goudet2018estimate}, anthropology~\cite{peletz1995kinship}, and computer science~\cite{wang2023survey}. Given a pair of measurements, each extracted from an individual, KV seeks to determine whether the two individuals are biologically related\footnote{Although this hypothesis-testing setting assumes a binary ground truth (\texttt{same} vs.\ \texttt{different}), all humans can be considered distant relatives, sharing more than $99\%$ of their genome. For practical applications, including our work, a sufficiently large generational separation between two individuals is treated as a true negative case, yielding a tractable detection formulation.}.

\begin{figure}[t]
\centering
\includegraphics[scale=0.30]{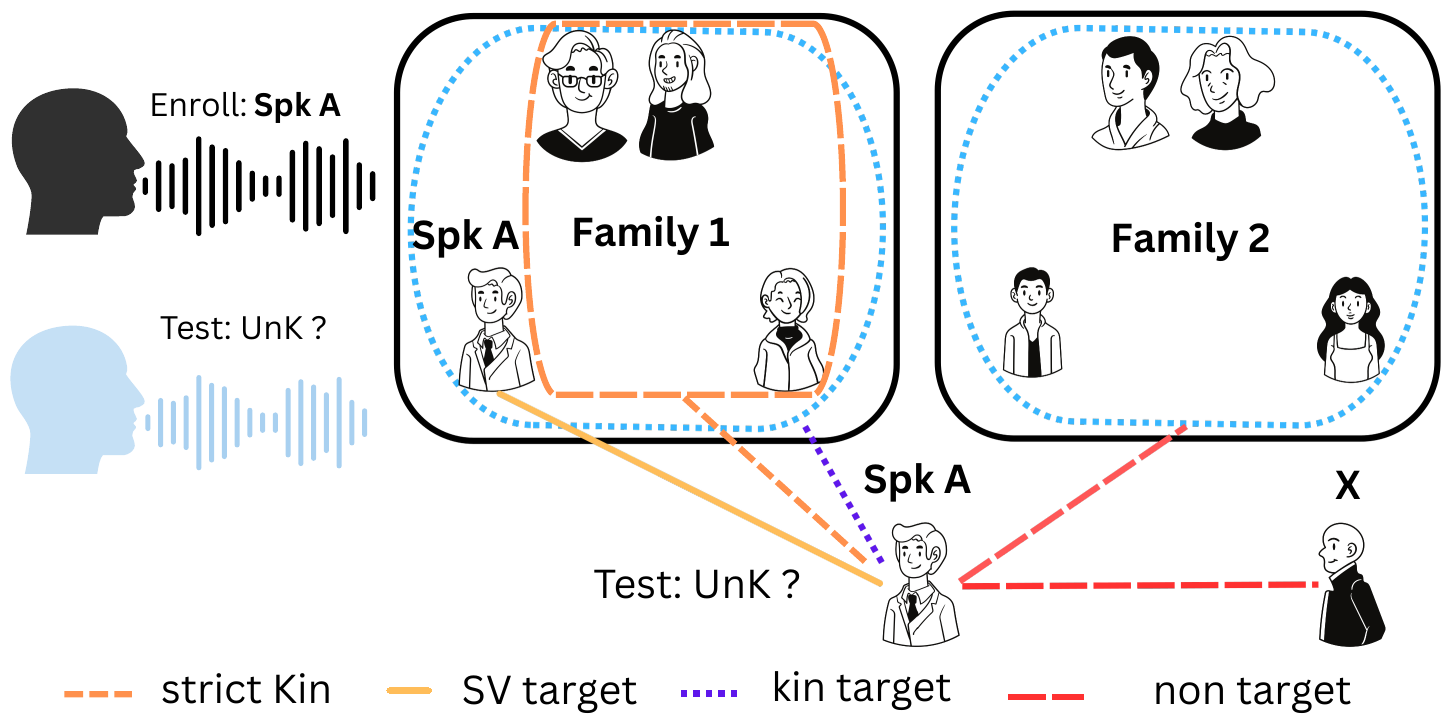}
\caption{Relation of kinship verification (KV) and speaker verification (SV). For \emph{Spk A} from \emph{Family 1}, KV treats trials with family members as \emph{kin-targets} and trials with speakers from other families as \emph{kin non-targets}. Self-comparisons (\emph{Spk A} vs. \emph{Spk A}) form \emph{SV target} trials, while comparisons with other family members form \emph{strict-kin} trials. In SV, only self-comparisons are targets. Thus, KV generalizes SV by extending the positive class from ``same speaker'' to ``same family''.
}
\label{fig:fam}
\end{figure}

Traditional approaches to KV rely on genetic analysis, with DNA profiling serving as the gold standard~\cite{goudet2018estimate}. While DNA provides an accurate and direct measure of genetic similarity, it is invasive, costly, time-consuming, requires specialized expertise and equipment, and is not universally applicable: not all crimes leave \emph{physical} traces, but may instead yield \emph{digital} evidence, such as video or voice recordings. These limitations have inspired research into non-invasive and more scalable KV approaches to complement DNA profiling.  Because related individuals share parts of their genetic makeup (\emph{genotype}), their \emph{phenotypes}—observable characteristics arising from the interaction of genotype and environment—can provide indirect evidence of kinship. Humans, for example, can reliably recognize facial kinship~\cite{kaminski2009human}, which has inspired automatic KV methods based on facial images and videos, including audio-visual approaches~\cite{wang2023survey,kefalas2023kan,robinson2021families}. Beyond visual cues, behavioral and linguistic traits have also been shown to contribute to the perception of kinship~\cite{Rykova2018-perception-and-acoustic-similarity,ball2018language,keen2014language}.


We focus on voice-based KV, a topic that has received only little attention. The human voice provides a distinctive modality for kinship analysis, reflecting both physiological and behavioral traits. As for the former, voiced speech is generated by airflow from the lungs passing through the trachea, causing the vocal folds to oscillate. The resulting quasi-periodic airstream is then shaped by the supralaryngeal vocal tract---including the pharyngeal, oral, and nasal cavities---and modulated by the articulators, such as the velum, jaw, tongue, and lips~\cite{quatieri2002discrete}. Because the anatomy of the respiratory system, vocal folds, and vocal tract is partly genetically determined, related individuals can be expected to exhibit similarities in their speech waveforms~\cite{nolanphonetic1983}. 
An extreme example is identical (monozygotic) twins, which have received
attention in forensic voice comparison~\cite{Kunzel2011AutomaticSR,NolanOh1996IdenticalTwins,
SABATIER2019-measurement-impact-twin,Alsalihi2024-effect-identical-twin-deep}
because of the potential confusability of family members' voices.


Beyond the morphology of the vocal production system and articulatory kinematics, which directly affect acoustic characteristics, speech is also shaped by behavior. This includes how sound units are organized into words and sentences, reflecting linguistic habits, social 
environment, and cognitive patterns~\cite{peletz1995kinship,ball2018language}. Anthropological studies suggest that shared linguistic traits among family
members are adaptive and may diminish in new socio-cultural
environments~\cite{william2001principles}.
In summary, the human voice can convey kinship cues through both acoustic properties and language use. 


Despite its relevance to both basic science and applications, voice-based KV
has received remarkably little attention. We focus on the
computational KV task, with the limited prior literature reviewed in Section~\ref{sec:related-work}. The prior studies either rely on small datasets or use evaluation protocols that may not accurately reflect the KV task. For instance, recent studies focus on closed-set 
protocols in which the same families appear in both training and evaluation data. 
Such an evaluation design does not fully capture \textbf{open-world}
verification, where systems must generalize to unseen speakers and families
without retraining. Although this setting is standard in speaker verification
(SV) benchmarking~\cite{Greenberg2020-two-decades,Doddington1985Speaker-Proc-IEEE}, it remains unaddressed in KV. In addition, without systematic control of confounders, reported performance may partly reflect
demographic, age-related, recording-quality, or dataset-specific shortcuts
rather than genuine kinship-related similarity.

We summarize our contributions as follows. \textbf{First}, we formulate voice-based KV as an open-world pairwise detection task, using family-disjoint splits to evaluate generalization to unseen families. \textbf{Second}, we establish a unified view of KV and SV, where SV is a special case of KV and strict-kin trials isolate familial similarity beyond speaker identity. \textbf{Third}, we introduce a curated KAN-AV evaluation protocol with matched trial construction to reduce demographic, age-, quality-, and noise-related confounding. \textbf{Fourth}, we benchmark modern speaker embeddings (ECAPA-TDNN~\cite{ecapadesplanques2020}, WavLM-ECAPA~\cite{wavlm}, and ReDimNet~\cite{yakovlev24_interspeech-REDIMNET}) 
with zero-shot and trainable KV back-ends to assess how well they encode familial cues. \textbf{Fifth}, we propose a lightweight asymmetric affine projection backend to better handle age- and gender-related variation. Together, these contributions provide a foundation for open-world voice-based kinship verification, covering task definition, evaluation design, and baseline detection methods.

\section{Related work}\label{sec:related-work}


\begin{table*}[t]
\centering
\caption{Studies on speech-based KV arranged chronologically. 
Dataset descriptions: In-house (12 families, Sanskrit recitations);  
TALKIN-Family (siblings, parent–child, grandparent–grandchild); KAN-AV (celebrity speech–video pairs).}
\label{tab:kinship_speech}
\begin{tabular}{p{0.8cm} p{2.2cm} p{3.6 cm}@{\hskip 20pt} p{9.4cm}}
\toprule
\textbf{Study} & \textbf{Dataset} & \textbf{Method} & \textbf{Key Findings} \\
\midrule
~\cite{padmini2017identification} & In-house & MFCC, VQ-LBG & Higher spectral correlation observed among same-gender Kin relatives; correlation weakens with increasing age differences and across genders. \\
~\cite{wu2019audio} & TALKIN-Family & I-vector, GMM-UBM, ResNet-50, cosine & Speech features provide complementary information to visual cues, enhancing kinship verification performance. \\
~\cite{kefalas2023kan} & KAN-AV & I-vector, x-vector, ResNet-50, triplet net & X-vectors capture kinship traits more effectively than i-vectors; audio cues supplement video, but performance declines for cross-gender pairs. \\
~\cite{wu2022audio} & TALKIN-Family & I-vector, pyannote, ResNet-50 & Incorporating speech improves detection accuracy, though performance decreases with greater genealogical distance and age differences. \\
~\cite{sun2024audio} & KAN-AV & Cycle-GAN (age conversion), i-vector, x-vector, triplet net & Age-conversion techniques can improve kinship recognition, but results are inconsistent across gender combinations. \\
\bottomrule
\end{tabular}
\end{table*}
\subsection{Kinship Recognition Using Voice}\label{subsec:kinship-from-voice-review}
A summary of related studies on voice-based kinship recognition is provided in Table~\ref{tab:kinship_speech}. Early work relied on handcrafted acoustic features, such as Mel-frequency cepstral coefficients (MFCCs) combined with vector quantization, on small in-house datasets. These studies reported stronger spectral correlations for same-gender kin pairs, with the effect weakening for larger age differences and across-gender pairs~\cite{padmini2017identification}.

Subsequent work introduced larger multimodal datasets and stronger speech
representations. Using the TALKIN-Family dataset~\cite{wu2019audio,wu2022audio}, the authors found that speech provides information complementary to visual
cues for kinship analysis, while performance degraded with increasing
genealogical distance and age difference. They considered Gaussian mixture model -- universal background model (GMM-UBM) and i-vector~\cite{dehak2010front} systems, together with pretrained embeddings such as ResNet-50~\cite{chung18b_interspeech} and PyanNet~\cite{bredin2020pyannote}, and was typically scored using cosine similarity. On the larger and more recent KAN-AV dataset introduced in~\cite{kefalas2023kan}, the authors evaluated i-vectors, x-vectors~\cite{snyder2018x}, and ResNet-50 features. The x-vector embeddings combined with a triplet-network backend provided the strongest performance among the tested speech representations, although results deteriorated for cross-gender relations. More recently,~\cite{sun2024audio} used CycleGAN-based~\cite{kaneko20_interspeech} age conversion together with i-vector, x-vector, and triplet-network modeling to improve performance on KAN-AV.

Although these studies have demonstrated that speech carries useful kinship-related information, their evaluation settings have certain shortcomings. In particular, recent works on the KAN-AV dataset~\cite{kefalas2023kan,sun2024audio} adopted triplet-based learning protocols in which anchor-positive pairs were formed from related speakers and negative samples were drawn from unrelated speakers, while the same kin-labelled subjects remained present across training and evaluation splits at different ages. Evaluation was then reported by classifying positive and negative pairs in a verification-like manner.  Consequently, these protocols do not fully reflect the open-world evaluation paradigm commonly used in SV research~\cite{nist2024sreplan}. 
This distinction is important because overlap of speakers or families across splits can encourage \emph{memorization} of subject- or family-specific characteristics rather than learning transferable kinship cues. 
Moreover, we identify various potential confounders in the KAN-AV data (detailed in Section~\ref{sec:data}). Lack of control for confounders risks bias in reported performance estimates.



\subsection{Impact of Kinship to Automatic Speaker Verification}

We are all trivially related to ourselves, sharing 100\% of our genes and behavioural traits. SV~\cite{Doddington1985Speaker-Proc-IEEE,Reynolds1995-speaker-identification-and-verification-GMM} is designed to detect this special form of kinship by carrying speaker comparison on a pair of utterances. The strong performance of modern SV models ~\cite{ecapadesplanques2020,yakovlev24_interspeech-REDIMNET} on challenging telephony and found-speech data~\cite{chung18b_interspeech,Greenberg2020-two-decades} has made SV practical in applications such as forensic voice comparison (FVC)~\cite{Rose2006-technical-forensic-speaker-recognition,MORRISON2021-consensus}. Progress in robust speaker feature extraction provides inspiration and initial optimism in addressing the broader class of KV tasks.

Although average SV performance has improved steadily over decades of research, it is well known that speaker comparison can fail under certain conditions. One such condition includes voice comparison of close family members, especially same-gender siblings. Their voices may be highly confusable, as phenotypic similarities, shared gender, and often similar age range jointly contribute to increased acoustic similarity. This has practical repercussions for FVC, for instance in composing the so-called \emph{relevant population}~\cite{HUGHES2015-relevant-population,MORRISON2021-consensus} used 
for statistical referencing purposes. 
For example, the baseline level of voice similarity appropriate for distinguishing between two brothers of the same age differs from that for comparing two unrelated males of the same age.

In SV performance estimation, the degree of family relatedness is an important evaluation-design factor, as our experiments also indicate (the curious reader may preview this effect in Fig.~\ref{fig:SD_KDE_Sub_SV}). Nevertheless, this factor has rarely been addressed beyond the forensics community. Aggregate performance measures computed over millions of speaker comparisons, such as in the NIST SRE~\cite{Greenberg2020-two-decades} and VoxCeleb~\cite{chung18b_interspeech} benchmarks, provide a useful summary of overall SV performance level. However, evaluations that explicitly account for family effects can offer more informative diagnostics for voice pairs that are likely to be highly confusable. Identical (\emph{monozygotic}, MZ) and non-identical (\emph{dizygotic}, DZ)
twins~\cite{NolanOh1996IdenticalTwins,SanSegundo2015-automatic-speaker-rec-Spanish-siblings,
Kunzel2011AutomaticSR,SANSEGUNDO2017-Euclidean}, as well as
siblings~\cite{Homayounpour1995-voice-discrimination-of-twins-and-siblings,
SanSegundo2015-automatic-speaker-rec-Spanish-siblings}, are examples of
speaker pairs whose voices may be confusable, particularly when they share the same gender and approximate age. This hypothesis has been confirmed using a variety of systems, including early neural-network and GMM approaches~\cite{Homayounpour1995-voice-discrimination-of-twins-and-siblings,ARIYAEEINIA2008-test-of-effectiveness-identical-twins}, distance-based methods~\cite{SANSEGUNDO2017-Euclidean}, i-vector embeddings~\cite{SABATIER2019-measurement-impact-twin}, and deep speaker embeddings~\cite{Alsalihi2024-effect-identical-twin-deep}. In summary, even if rarely analyzed, SV performance is remarkably influenced by family relatedness.

\section{Speaker and kin verification: a unified view}


\subsection{Definitions}

Let $\mathcal{X} = \{x_i\}$ denote a set of speech recordings, each assumed to contain speech from exactly one individual. To each recording $x_i$ we associate a pair of categorical ground-truth labels $\vec{c}_i=(c^{\spk}_i,c^{\famlab}_i)$. The first label, $c^{\spk}_i \in \{1, \dots, S\}$, uniquely identifies the speaker, while the second, $c^{\famlab}_i \in \{0, 1, \dots, F\}$, identifies the biological family to which the speaker belongs. The special value $c^{\famlab}_i\!=\!0$ is reserved to denote an \emph{out-of-family} individual, that is, a speaker not associated with any of the $F$ known families.

The two labels are related through a hierarchical relation: the speakers are \emph{nested} within the families. The two labels are therefore related by a deterministic mapping that assigns every speaker to exactly one family (or to the ``unknown family''). Because every speaker belongs to at most one family, the speaker partition forms a
\emph{refinement} of the family partition, i.e. $c^{\spk}_i\!=\! c^{\spk}_j\;\Rightarrow\;c^{\famlab}_i\!=\!c^{\famlab}_j$ holds (but not vice versa).

Whereas multi-class treatment of speaker and kinship recognition are possible, we purposefully focus on the more elementary two-class (or verification) task. It consists of either accepting or rejecting a claimed identity relationship between a pair of utterances, which can be either a claim on speaker or family identity. We define a \emph{trial} as a pair of recordings $t\!=\!(x_i, x_j) \in \mathcal{X}\!\times\!\mathcal{X}$, and let $\mathcal{T}$ to denote the set of all trials. Each trial $t \in \mathcal{T}$ has two well-defined binary \emph{trial keys}, corresponding to the two tasks:
\begin{align}
  y^{\spk}_t &=
  \begin{cases}
    1, & \text{if } c^{\spk}_i = c^{\spk}_j\;\;\;\;\text{(speaker target)}\\
    0, & \text{otherwise}\;\;\;\;\;\;\;\;\;\;\text{(speaker nontarget)},
  \end{cases}
  \label{eq:key_spk} \\[6pt]
  y^{\famlab}_t &=
  \begin{cases}
    1, & \text{if } c^{\famlab}_i = c^{\famlab}_j \neq 0\;\;\;\;\text{(kin target)}\\
    0, & \text{otherwise}\;\;\;\;\;\;\;\;\;\;\;\;\;\;\;\;\;\;\text{(kin nontarget)}
  \end{cases}
  \label{eq:key_fam}
\end{align}
The condition $c^{\famlab}_i\!=\!c^{\famlab}_j\!\neq\!0$
in kin label ensures that two out-of-family individuals will not be treated as kin. Following standard SV benchmarking nomenclature~\cite{nist2024sreplan,Greenberg2020-two-decades}, we refer to any trial with
$y_t^\bullet = 1$ as a \emph{positive} or \emph{target} trial and those with $y_t^\bullet = 0$
as a \emph{negative} or \emph{non-target} trials. We add the determiner 'speaker' or 'kin' when necessary.

With the above notations, \textbf{speaker
verification} (or speaker detection) requires adjudicating whether $y_t^\text{spk}\!=\!1$ (speaker null hypothesis) or not (speaker alternative hypothesis). In a similar vein, \textbf{kinship verification} is the task of determining whether $y_t^\famlab=1$ (kin null hypothesis) or not (kin alternative hypothesis). Both tasks involve two steps: computing a \emph{detection score}, where larger values indicate stronger support for the null hypothesis, and thresholding it to make a binary decision.

\begin{table}[ht]
\centering
\caption{Trial partition induced by nesting of speaker and family labels. The cell marked ``$\emptyset$'' is structurally empty.}
\label{tab:partition}
\small
\setlength{\tabcolsep}{8pt} 
\begin{tabular}{@{}lll@{}}
\toprule
 & \multicolumn{1}{l}{$y^{\famlab}_t=0$} & \multicolumn{1}{l}{$y^{\famlab}_t=1$} \\
\midrule
$y^{\spk}_t=0$ 
  & \begin{tabular}[l]{@{}l@{}}Different family\\ or out-of-family\end{tabular} 
  & \begin{tabular}[l]{@{}l@{}}Different speaker,\\ same family (\emph{strict kin})\end{tabular} \\
\addlinespace[0.8ex]
$y^{\spk}_t=1$ 
  & $\emptyset$
  & Same speaker \\
\bottomrule
\end{tabular}
\end{table}


\subsection{Interaction of Speaker and Kin Detection Tasks}\label{subsec:speaker-kin-interaction}

The nesting of speakers within families implies a strict logical
relation between the two trial keys: $y^{\spk}_t\!=\!1
\;\Rightarrow\; y^{\famlab}_t\!=\!1$. Equivalently, the target set of speaker verification is a proper
subset of the target set of kinship verification. Consequently, every
trial falls into exactly one of three mutually exclusive classes
(see Table~\ref{tab:partition}):
\begin{enumerate}
  \item \textbf{Same speaker}
        ($y^{\spk}_t\!=\!1 \land y^{\famlab}_t\!=\!1$): Both recordings originate
        from the same individual (and hence also from the same family).

  \item \textbf{Strict kin}
        ($y^{\spk}_t\!=\!0 \land y^{\famlab}_t\!=\!1$): The utterances are
        produced by two different individuals who belong to the same known
        family. These trials are targets for kinship verification but
        non-targets for speaker verification.

  \item \textbf{Unrelated}
        ($y^{\spk}_t\!=\!0 \land y^{\famlab}_t\!=\!0$): The utterances come
        from speakers in different families, or at least one speaker
        is out-of-family. These are non-targets for both tasks.
\end{enumerate}

In SV, class~1 forms the target set and classes~2 and~3 are non-targets; in
KV, classes~1 and~2 are targets and class~3 is non-target. 
The most informative KV trials are strict-kin pairs, since
they require detecting shared familial voice characteristics without relying
on speaker identity; same-speaker pairs (class~1), although technically valid kin targets, can be resolved by SV alone and do not probe pure kinship-specific information. Accordingly, we evaluate KV under two conditions: \emph{overall} KV (\textbf{KV}), which includes speaker-target trials, and \emph{strict} KV (\textbf{KV$^{*}$}), which excludes them. 




Note that familial voice similarity plays opposite roles in KV and SV. Biologically or environmentally close speakers, due to shared genetics, anatomy, accent, or household exposure, may sound more similar. In KV, this similarity is \emph{useful}: close relatives form easier positive trials. In SV, it is \emph{detrimental}: close relatives form harder non-target trials, as the system must reject potentially similar voices. Monozygotic twins are the extreme case, but the effect extends to siblings, parent--child pairs, and other close relatives. Thus, the familial composition of the evaluation set can substantially affect measured performance in both tasks, but in opposite directions.

\section{On The Design of Evaluation Trials}
\label{Sec:design}

Having defined the SV and KV tasks, we now turn to the design of their performance evaluation. This requires organising a corpus of recordings $\mathcal{X} = \{x_i\}$ into a set of pairwise comparisons (evaluation trials), each associated with a binary ground-truth following \eqref{eq:key_spk} or \eqref{eq:key_fam}. 
As a modest contribution toward establishing sound evaluation practices for KV, a relatively new task in speech research, we briefly review the theoretical aspects relevant to evaluation design.

A seemingly natural approach is to generate trials by exhaustively
pairing all (nonidentical) utterances. This exposes a system to diverse trial configurations, mimicking uncontrolled ``in-the-wild'' data,
a reasoning that has implicitly guided many speech dataset
designs. Beyond the obvious computational
drawback\footnote{$\mathcal{O}(N^2)$ trials in the number of
recordings, $N$.}, uncritical cross-pairing can introduce
systematic and unnoticed biases into performance evaluation. The root issue is that a waveform $x_i$ is not generated
solely from the speaker or family identity, but is also causally
influenced by population demographics and recording conditions, many of which are shared across subsets of recordings. When trials are formed without regard to this structure, the trial set entangles the task-relevant semantic relation with auxiliary factors. Evaluation metrics may then reflect the exploitation of incidental correlations (\emph{shortcuts}%
~\cite{geirhos2020shortcut}) rather than the system's ability to
detect the intended identity relation.

\subsection{Latent Speaker Attributes with Hierarchical Structure}

Speech waveforms depend on physiological and behavioral
traits of speakers. For waveform $x_i$, we represent them by a triplet of latent variables $\boldsymbol{h}_i=\bigl(\boldsymbol{h}_i^{\famlab}, \boldsymbol{h}_i^{\spk},
\boldsymbol{h}_i^{\aux}\bigr)$
where $\boldsymbol{h}_i^{\famlab}$ captures familial or genetic traits,
$\boldsymbol{h}_i^{\spk}$ represents speaker-specific deviations from those
familial traits, and $\boldsymbol{h}_i^{\aux}$ denotes additional factors such
as speaker's age, gender, or room acoustics. 
The ground-truth labels ($c_i^{\spk}$ and $c_i^{\famlab}$) are not primitive causes but symbolic summaries of the latent family and speaker factors. This direction of causation (from latent attributes to observed class labels) follows the
physical data-generating process, commonly used in causal modeling
\cite{pearl2009causality}.

For a trial $t\!=\!(x_i,x_j)$, the joint latent configuration
$(\boldsymbol{h}_i,\boldsymbol{h}_j)$ induces the trial-level variables
$\bigl(y_t^{\spk},y_t^{\famlab},Z_t\bigr)=
g(\boldsymbol{h}_i,\boldsymbol{h}_j)$,
where $g(\cdot)$ is a deterministic map, and $Z_t$ encapsulates the combined auxiliary factors from the two recordings. This mapping should be understood in a
purely definitional sense: it reflects how the trial labels (ground truth) and auxiliary descriptors are \emph{assigned} (by a corpus designer), rather than computed by an algorithm. 
Unlike in \emph{interventional} settings (e.g., treatment effects in clinical studies), the trial labels are not manipulable variables in any physical sense. Nevertheless, the causal notion remains useful as a conceptual reference\footnote{To elaborate, let $s_t \in \mathbb{R}$ denote a detection score produced by a system. The \emph{interventional distribution} of $s_t$ is $P(s_t \mid \mathrm{do}(y_t^\bullet=y))$, where 'do' represents the intervention (manipulation of) the label variable. This does not describe a realizable operation on trial labels, but an \emph{idealized comparison} where the semantic relation
encoded by the trial key is the only factor allowed to vary, while all other
latent pairwise attributes---such as gender, age, and channel---are held constant. In practice, evaluation is necessarily \emph{observational}, yielding familiar conditional distributions of the form
$P(s_t \mid y_t^\bullet=y)$. The role of trial design is therefore not to implement an intervention on $y_t^\bullet$, but to control auxiliary pairwise variables so that conditioning on $y_t^\bullet$ approximates the idealized interventional comparison. This causal perspective motivates the use of controlled or matched trial subsets.}.

\subsection{Detection scores as mixture distributions}

Let $y_t^\bullet$, where $\bullet \in \{\mathrm{spk}, \mathrm{fam}\}$, denote the task-specific trial key. The class-conditional score distribution can be written as
\begin{equation}
\begin{aligned}
P(s_t \mid y_t^\bullet=y)
&=
\int P(s_t \mid y_t^\bullet=y, Z_t=z)\,
     p(z \mid y_t^\bullet=y)\, dz \\
&=
\mathbb{E}_{Z_t \mid y_t^\bullet=y}
\!\left[ P(s_t \mid y_t^\bullet=y, Z_t) \right].
\end{aligned}
\label{eq:mixture_general}
\end{equation}
Equation~\eqref{eq:mixture_general} highlights that evaluation is performed by
averaging score distributions over subsets of trials that share the same task
label but differ in auxiliary pairwise attributes. Hence, the class-conditional score distributions (and performance metrics derived from them) depend not only on the system behavior within each subset, but also on the prevalence of these subsets in the evaluation protocol. This is undesirable because the relative prevalence of the subsets is a consequence of dataset collection and trial construction, rather than a fundamental property of the verification task. As a result, reported performance metrics may change simply due to differences in trial composition, even when the underlying system and data remain the same. This, in turn, renders performance metrics less comparable across studies.

A simple example may be helpful to illustrate the issue. Consider speaker verification with a binary auxiliary variable $Z_t \in \{0,1\}$ that encodes whether the two speakers have the same gender (\emph{or} language, \emph{or} acoustic environment, etc.). In this case, the non-target score distribution can be written as
\begin{equation}
\begin{aligned}
P(s_t \mid y_t^{\spk}=0)
&=
\rho_0 \, P(s_t \mid y_t^{\spk}=0, Z_t=0) \\
&\quad +
(1-\rho_0)\, P(s_t \mid y_t^{\spk}=0, Z_t=1),
\end{aligned}
\label{eq:mixture_gender}
\end{equation}
where $\rho_0 = P(Z_t=0 \mid y_t^{\spk}=0)$ denotes the proportion of non-target
trials with different-gender speakers.

Equation~\eqref{eq:mixture_gender} has direct consequences for
performance measurement. Let $\tau$ denote a detection threshold.
The false-alarm (FA)
rate $P_{\text{fa}}(\tau)$ is the conditional probability that the
score exceeds $\tau$ given a non-target trial ($y_t^{\spk}=0$), and
the miss rate $P_{\text{miss}}(\tau)$ is the conditional probability
that the score falls below $\tau$ given a target trial
($y_t^{\spk}=1$). Substituting Eq.~\eqref{eq:mixture_gender} into
these definitions yields the subgroup decompositions
\begin{equation}
\begin{aligned}
P_{\text{fa}}(\tau)
&=
\rho_0 \, P_{\text{fa}}^{(0)}(\tau)
+
(1-\rho_0)\, P_{\text{fa}}^{(1)}(\tau), \\
P_{\text{miss}}(\tau)
&=
\rho_1 \, P_{\text{miss}}^{(0)}(\tau)
+
(1-\rho_1)\, P_{\text{miss}}^{(1)}(\tau),
\end{aligned}
\label{eq:pfa_pmiss-mixture}
\end{equation}
where $P_{\text{fa}}^{(z)}(\tau)$ and $P_{\text{miss}}^{(z)}(\tau)$
denote the subgroup-specific FA and miss rates conditioned
on $Z_t\!=\!z$, with $z \in \{0,1\}$.


Equation~\eqref{eq:pfa_pmiss-mixture} makes explicit that
both error rates are weighted averages of the respective subgroup-specific error rate, with weights determined by the empirical prevalence of the auxiliary variable within the two trial classes. 
Consequently, changes in trial composition alone can alter the operating characteristics of a system, even when the underlying score distributions within each subgroup remain unchanged.

\subsection{Hard Trials and Nonparametric Standardization}\label{subsec:non-parametric-std}

To address the issues noted above, it is necessary to reduce the influence of auxiliary pairwise attributes on the evaluation outcome. Two complementary strategies are commonly employed. The first is the selection of restricted or ``hard'' trial subsets that fix certain auxiliary variables to challenging or informative values---an example from the speaker verification literature~\cite{Greenberg2020-two-decades} is exclusion of opposite-gender or mismatched-language nontargets. Selecting hard trials corresponds to conditioning on specific values of $Z_t$ in
\eqref{eq:mixture_general}.  

The second is \emph{non-parametric standardization}~\cite{hernan2020causal}, in which trial sets are constructed so that auxiliary attributes follow comparable
empirical distributions across trial classes. This is analogous to
the use of \emph{control groups} in experimental study design:
just as a control group is matched to the treatment group on
potentially confounding covariates so that the treatment effect can
be isolated, here the non-target (or non-kin) trials are matched to
the target (or kin) trials on auxiliary attributes so that the
measured detection performance reflects the intended identity
relation rather than incidental differences in the trial population. The standardized distribution provide a closer estimate of the system’s
\emph{task-relevant discriminative capability} than metrics computed under an
uncontrolled observational evaluation.

Age-difference matching, particularly relevant to kinship detection where large age difference between parents and their children are expected \emph{a priori}, provides a concrete illustration.
Let $Z_t = |\text{age}_i - \text{age}_j|$ denote the absolute difference of ages of the speakers in the recordings involved in trial $t$. By constructing evaluation trials such that
\begin{equation}
p(Z_t \mid y_t^\bullet=1)
=
p(Z_t \mid y_t^\bullet=0),
\label{eq:standardization}
\end{equation}
ensures that both class-conditional score distributions in
Eq.~\eqref{eq:mixture_general} are averaged with respect to the same empirical
distribution over the auxiliary variable $Z_t$.
As a result, any differences between
$P(s_t \mid y_t^\bullet\!=\!1)$ and $P(s_t \mid y_t^\bullet\!=\!0)$ arise solely from
differences in the conditional score distributions
$P(s_t \mid y_t^\bullet=y, Z_t=z)$, rather than from differences in how auxiliary
trial configurations are weighted.

\section{Revised KAN-AV Evaluation Protocol}

\begin{figure*}[t]
\centering
\includegraphics[height= 70pt,width=420pt]{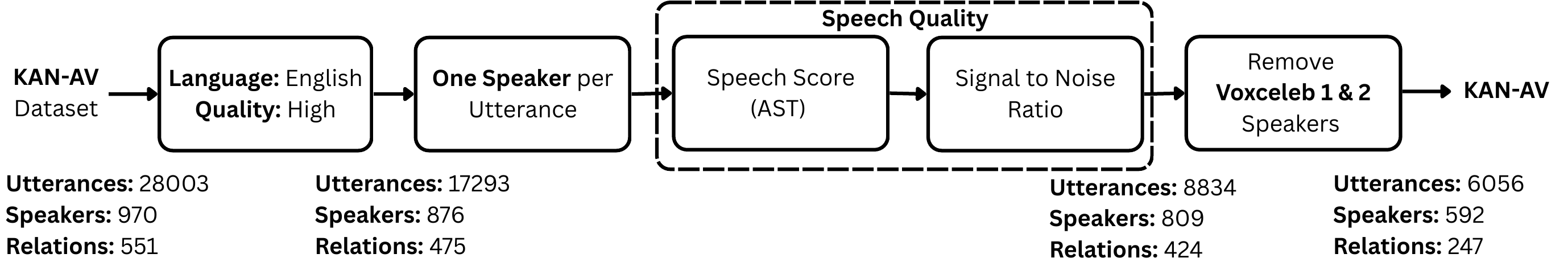}
  \caption{Pipeline for constructing the curated speech subset from KAN-AV. The filtering stages include language and quality selection, single-speaker filtering, speech-score and signal-to-noise screening, and removal of recordings or speakers likely to introduce nuisance variability.}
\label{fig1:Datapreparation_Pipeline}
\end{figure*}

\label{sec:data}
\subsection{Data Selection}
The original KAN-AV dataset~\cite{kefalas2023kan} contains $28,003$ recordings (crawled from YouTube) from $970$ unique speakers. It contains $551$ unique 
pairwise kin relations. 
All speakers are public figures, such as actors, musicians, and politicians, with recordings spanning various periods of their lives. The kin relations in the original KAN-AV dataset were derived using metadata available in Wikipedia and the IMDb website~\cite{georgopoulos2020investigating}.


As the largest dataset of its kind, KAN-AV represents a valuable contribution to kinship research. Nonetheless, in its raw form, it is not directly usable for our study. Our initial investigation indicated that, as a form of \emph{found data}, the dataset includes characteristics that may introduce nuisance variability unrelated to kinship. To mitigate these effects, we adopt a multi-step data-filtering pipeline (Fig.~\ref{fig1:Datapreparation_Pipeline}) to exclude recordings that could bias the analysis or obscure kinship-related patterns.

First, to retain phonotactic and lexical consistency, we use the language labels in the original KAN-AV metadata to retain only English utterances. We also retain only recordings labeled as \emph{high quality} in the metadata. Despite this initial filtering, we identified several recordings containing strong background sounds (e.g., music and clapping) and multiple speakers. We therefore used the \texttt{Pyannote} toolkit~\cite{bredin2023pyannote} to first estimate the number of speakers in each recording, retaining only single-speaker recordings. We then employed the \emph{audio spectrogram transformer} (AST) model~\cite{AST21b_interspeech} to detect speech-only events and the \emph{waveform amplitude distribution analysis} (WADA) method~\cite{kim08e_interspeech} to estimate the signal-to-noise ratio (SNR). We retained recordings for which the speech class probability was $\geq\!0.65$ and SNR $\geq\!0$ dB. These thresholds were set empirically to balance segment quality and dataset size. 

Finally, we removed all speakers overlapping with VoxCeleb~\cite{chung18b_interspeech}. This enables us to use speaker-embedding extractors pretrained on VoxCeleb1 and VoxCeleb2 without speaker leakage. The final curated subset of KAN-AV, comprising approximately $\sim\!20\%$ of the original dataset, contains $6{,}056$ utterances from $592$ speakers covering $247$ kinship relations.

\begin{figure*}[t]
\centering
\includegraphics[height= 85pt,width=480pt]{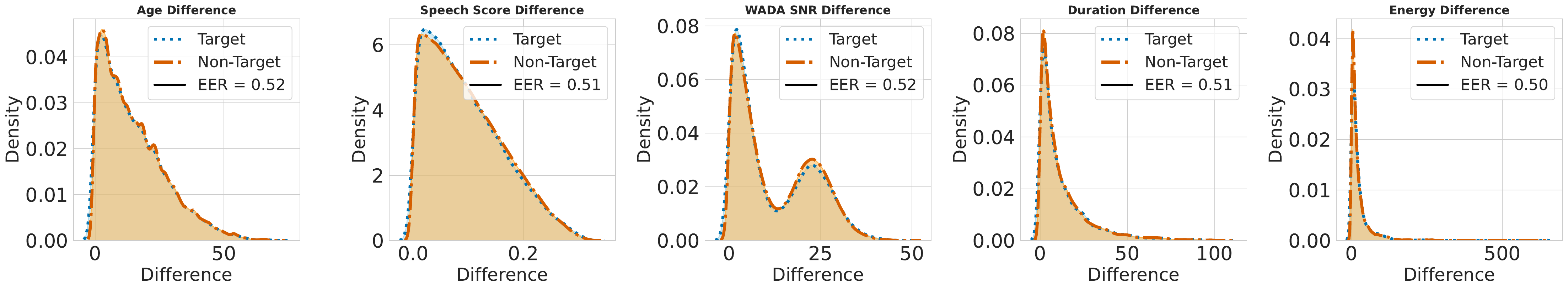}
  \caption{Distributions of five trial-level confounding factors for kin-target and kin non-target pairs in the test set.}
\label{fig3:kde_diff}
\end{figure*}

\begin{figure}[t!]
\centering
\includegraphics[scale=0.35]{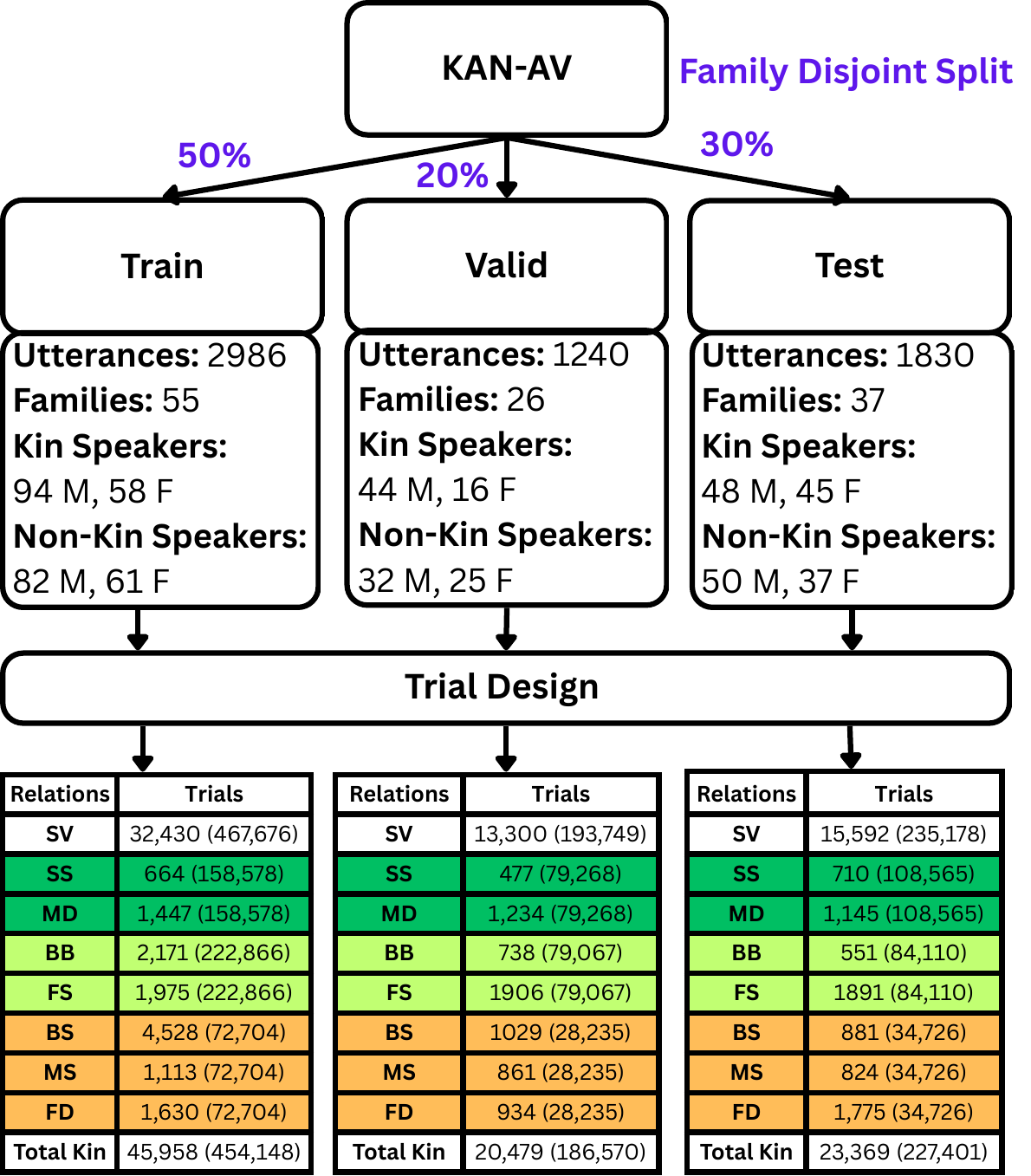}
  \caption{Dataset split and trial design. Figure reports the number of utterances, families, speakers, and relation-specific trials in the train, validation, and test, using a family-disjoint partition and gender-aware trial construction. SS: Sister--Sister, MD: Mother--Daughter, BB: Brother--Brother, FS: Father--Son, BS: Brother--Sister, MS: Mother--Son, and FD: Father--Daughter.}
\label{fig2:Trail_Pipeline}
\end{figure}


\subsection{Evaluation Protocol Design}

The resulting subset of KAN-AV serves as the basis for the evaluation protocol design. Among the $592$ retained individuals, $305$ belong to one of $118$ unique families, while the remaining $287$ are \emph{out-of-family} speakers ($c^{\famlab}_i\!=\!0$, with no known family ties to any other speaker). Following the \emph{open-set} protocol design philosophy (see Subsection~\ref{subsec:kinship-from-voice-review}), we partition the former group into training, validation, and test splits in a $50\!:\!20\!:\!30$ ratio using a \emph{family-disjoint} design---that is, data from any given family appear in exactly one of the three splits. In addition, we aim to keep the male-to-female ratio in each split as close as possible to that of the selected KAN-AV subset. Because speakers are nested within families, speaker identities are also disjoint across the three splits. The out-of-family speaker set is partitioned using the same ratios.


The two resulting sets of training, validation, and test data are then merged to define the pairwise comparisons (trials) used for both the SV and KV tasks. For both tasks, the corresponding trial key (ground truth) is binary, indicating whether the two utterances in a trial pair share the same speaker identity or the same family identity. With reference to Subsection~\ref{subsec:speaker-kin-interaction} and Fig.~\ref{fig:fam}, however, the negative (non-target) class differs between the SV and KV tasks.

From the complete pool of kin non-target pairs, we construct an evaluation set that is non-parametrically standardized (see Subsection~\ref{subsec:non-parametric-std}) with respect to two covariates: gender composition and age difference. The standardization is performed on a per-target-pair basis: for each kin-target pair, we sample ten kin non-target pairs from the pool that exhibit the same gender combination and the same age difference. This matching procedure ensures that the empirical distributions of these two covariates are approximately identical (by construction) across the target and non-target trial sets.  Fig.~\ref{fig2:Trail_Pipeline} illustrates the resulting trial statistics and the trial generation pipeline.

Beyond demographic matching, the protocol should also diagnose
potential \emph{shortcut} cues in the constructed trials
\cite{sahidullah2025shortcut,geirhos2020shortcut}. Otherwise, a
detector may exploit auxiliary similarity, such as age or recording
quality, rather than kinship-related voice cues. We therefore compare
kin-target and kin-nontarget trials using five trial-level confounders,
computed as absolute pairwise differences: speaker age, AST speech
score~\cite{AST21b_interspeech}, WADA-SNR~\cite{kim08e_interspeech},
rVAD-based active speech duration~\cite{tan2020rvad}, and the
active-speech-to-total-energy ratio. These factors capture age-related
vocal similarity, speech salience, additive noise mismatch, usable
speech content, and the dominance of speech over nonspeech events,
respectively. As shown in Fig.~\ref{fig3:kde_diff}, the target and
non-target distributions are closely aligned after trial matching, with
EERs near chance level ($\approx 50\%$). Similar trends are observed
for the training and development splits, indicating that these nuisance
variables provide little class-discriminative information.

\section{Kinship Verification via Speaker Embeddings}\label{sec:kinship-verification}

Since speaker embeddings~\cite{wang2024overview,ecapadesplanques2020} can be assumed to encode both physiological (e.g., vocal tract length) and behavioral speaker traits (e.g., idiolect), we hypothesize that they also carry information indicative of broader kin relations beyond speaker identity. To investigate this systematically, we consider two complementary evaluation paradigms. In the first, a \emph{zero-shot} approach, the speaker embedding extractor is used as-is, and pairwise kin similarity is computed directly via cosine scoring. In the second, a \emph{trainable-backend} approach, kinship is modeled by a Siamese architecture~\cite{kefalas2023kan} trained on top of the embeddings. These two paradigms enable us to assess the extent to which speaker embeddings encode familial similarity and to establish a scalable kinship-verification pipeline applicable to both zero-shot and supervised settings.


\subsection{Zero-shot Kinship Verification}

Given a trial pair with embeddings ${\bf e}_1, {\bf e}_2 \in \mathbb{R}^d$, we compute their cosine similarity as $ s(\mathbf{e}_1, \mathbf{e}_2) = (\mathbf{e}_1^\top \mathbf{e}_2) / (\|\mathbf{e}_1\|_2 \, \|\mathbf{e}_2\|_2) $, where $^\top$ and $\|\cdot\|_2$ denote transpose and Euclidean norm, respectively. Higher cosine scores indicate stronger kin evidence.

\subsection{Trainable Backends}
\label{sec:trainable_backend}

In the trainable setting, we consider three network designs, the first two of which follow a conventional Siamese architecture. First, we implement a \emph{fully connected} architecture inspired by TripletNet-based KV methods~\cite{kefalas2023kan}. Second, given the limited number of available training trials~\cite{barz2020deep}, we employ a simple \emph{symmetric affine projection} that applies a shared affine transformation to both embeddings before cosine scoring. The model is trained under a constrained objective that regularizes the transformation toward an identity mapping, thereby preserving the cosine geometry and speaker-discriminative structure of the original embedding space while adapting it to KV. Third, to address the age and gender mismatches that frequently arise in KV, we introduce an  \emph{asymmetric affine projection} in which only one embedding is transformed while the other remains fixed (see Fig.~\ref{fig:AS_AP}). The intuitive notion is that the transformation encourages moving the age- and/or gender latent factors of one of the embeddings closer to the other one.

\subsubsection{Fully Connected Network (FCN)}

Inspired by~\cite{kefalas2023kan}, we adopt a three-layer fully connected network that maps a $d$-dimensional speaker embedding through two hidden layers of size $d/2$. Each linear transformation is followed by batch normalization and a nonlinear activation function to improve training stability and generalization. The transformation is
\begin{equation}
\begin{aligned}
{\bf z}_i = f_\theta({\bf e}_i) 
= \phi_3\Big(&\mathrm{BN}_3\big({\bf W}_3 \,
\phi_2(\mathrm{BN}_2({\bf W}_2 \, \\
&\phi_1(\mathrm{BN}_1({\bf W}_1 {\bf e}_i + {\bf b}_1)) + {\bf b}_2))
+ {\bf b}_3\big)\Big),
\end{aligned}
\end{equation}
where $i \in \{1,2\}$ indices the two embeddings in a trial, ${\bf W}_k$ and ${\bf b}_k$ denote the weight matrices and bias vectors, respectively, $\mathrm{BN}_k(\cdot)$ denotes batch normalization, and $\phi_k(\cdot)$ denotes a nonlinear activation function. In our experiments, we use the rectified linear unit (ReLU) as the activation function for all layers, i.e. $k \in \{1,2,3\}$.

The network parameters are optimized using a contrastive loss~\cite{hadsell2006dimensionality}, which encourages embeddings of kin-target pairs to be close in the transformed space while pushing embeddings of kin non-target pairs farther apart. The loss is defined as
\begin{equation}
\mathcal{L}_{c}
= y \, \|{\bf z}_1 - {\bf z}_2\|_2^2
+ (1-y)\,\max\!\left(0, m - \|{\bf z}_1 - {\bf z}_2\|_2\right)^2,
\end{equation}
where $y \in \{0,1\}$ indicates whether the pair is a kin-target pair ($y=1$) or a kin non-target pair ($y=0$), and $m$ denotes the margin parameter. Following~\cite{kefalas2023kan}, we further regularize the learned representations by adding an $L_2$ penalty on the transformed embeddings during training. The overall objective is
\begin{equation}
\mathcal{L}
= \mathcal{L}_{c}
+ \lambda_f \sum_{i=1}^{2} \|f_\theta({\bf e}_i)\|_2^2,
\end{equation}
where $\lambda_f$ controls the strength of the regularization and is empirically tuned using the validation set.

\subsubsection{Symmetric Affine Projection (S-AP)}
\label{sec:tr_bacend_S_AP}
This backend provides a lightweight trainable adaptation~\cite{barz2020deep} 
that aims to preserve the structure of the original speaker embedding space. In particular, we apply the same affine transformation to both embeddings in a trial and perform verification using cosine similarity in the transformed space. The transformation is regularized toward the identity mapping to prevent excessive deformation of the speaker-embedding space topology~\cite{kulis2013metric,davis2007information}, thereby reducing the risk of overfitting and preserving discriminative speaker information learned during large-scale pretraining~\cite{nagrani2020voxceleb}. Preserving this structure is particularly important in low-resource KV settings, where aggressive transformations may over-specialize to the training data and degrade generalization to unseen speakers and recording conditions. Moreover, maintaining the original embedding geometry helps retain robustness to nuisance variability already encoded by the pretrained speaker model.


Formally, each input embedding is transformed as
\begin{equation}
{\bf z}_i = f_\theta({\bf e}_i) = {\bf W}{\bf e}_i + {\bf b},
\end{equation}
where ${\bf W} \in \mathbb{R}^{d \times d}$ is a linear transformation matrix and ${\bf b} \in \mathbb{R}^{d}$ is a bias vector. The transformed embeddings are subsequently length-normalized and compared using cosine scoring. We optimize ${\bf W}$ and ${\bf b}$ using the combined objective
\begin{equation}
\mathcal{L} = \mathcal{L}_{\rm cosine} + \mathcal{L}_{\rm reg},
\end{equation}
where
\begin{equation}
\mathcal{L}_{\rm reg} = \lambda_w \|{\bf W} - {\bf I}\|_F^2 + \lambda_b \|{\bf b}\|_2^2.
\end{equation}
Here, ${\bf I}$ denotes the identity matrix and $\lambda_w$ and $\lambda_b$ are regularization coefficients that are empirically tuned using the validation set. The first term encourages ${\bf W}$ to remain close to the identity mapping, while the second term constrains ${\bf b}$ to avoid unnecessary global shifts of the embeddings that could distort cosine-based similarity relationships.

The cosine embedding loss is defined as
\begin{equation}
\mathcal{L}_{\rm cosine} =
\frac{1}{N} \sum_{i=1}^N
\begin{cases}
1 - s({\bf z}_1^{(i)}, {\bf z}_2^{(i)}), & y_i = 1, \\
\max\!\left(0, s({\bf z}_1^{(i)}, {\bf z}_2^{(i)}) - m\right), & y_i = -1,
\end{cases}
\end{equation}
where $y_i \in \{1,-1\}$ denotes the kin-target and kin non-target labels, respectively, and $m$ is the margin parameter.

\subsubsection{Asymmetric Affine Projection (AS-AP)}
\label{sec:tr_bacend_AS_AP}

\begin{figure}[t!]
\centering
\includegraphics[height= 70pt,width=190pt]{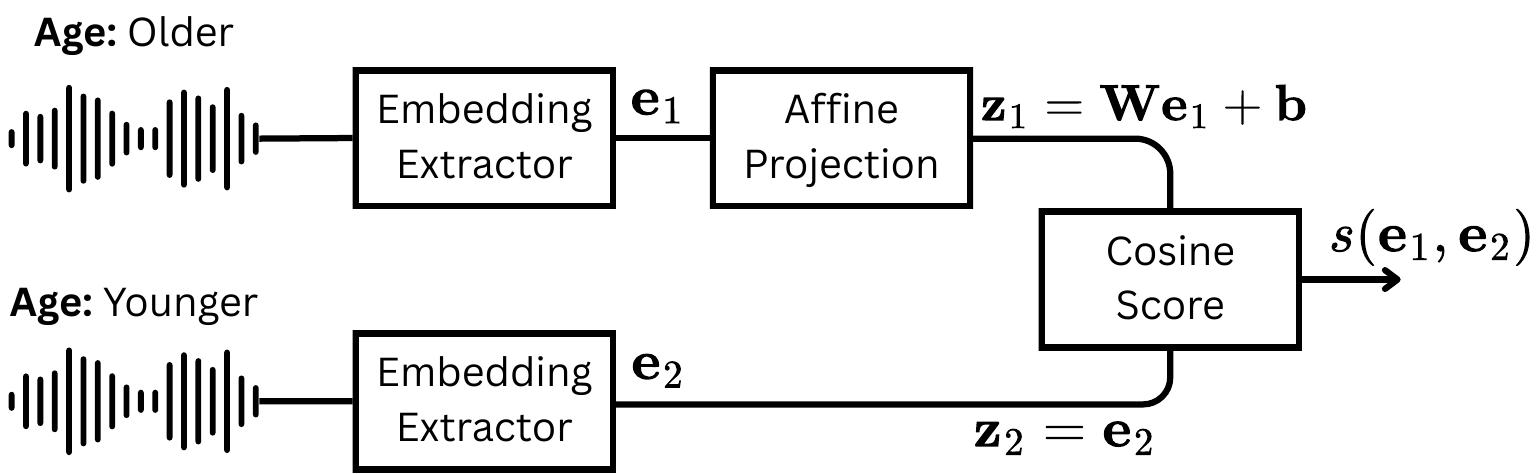}
  \caption{Proposed Asymmetric Affine Projection (AS-AP) trainable backend.}
\label{fig:AS_AP}
\end{figure}

In our proposed \textbf{asymmetric affine projection} backend (Fig.~\ref{fig:AS_AP}), only one embedding is transformed while the other remains fixed. This leads to a directional scoring formulation in which the embedding order matters. In our implementation, we impose a deterministic ordering during both training and evaluation, based on demographic metadata.  For \emph{age-constrained} projection, the older speaker's embedding is assigned to ${\bf e}_1$ and the younger speaker's to ${\bf e}_2$. For \emph{gender-constrained} projection, the male embedding is assigned to ${\bf e}_1$ in cross-gender trials. For \emph{joint age-and-gender-constrained} projection, we first apply age-based ordering and then enforce the gender-based rule for cross-gender pairs. This ordering removes directional ambiguity and encourages the model to learn systematic kin-related transformations associated with age and gender.



Formally, given an ordered trial pair ${\bf e}_1, {\bf e}_2 \in \mathbb{R}^{d}$, we define
\begin{equation}
\begin{cases}
{\bf z}_1 = f_\theta({\bf e}_1) = {\bf W}{\bf e}_1 + {\bf b}\\
{\bf z}_2 = {\bf e}_2,
\end{cases}
\end{equation}
where ${\bf W} \in \mathbb{R}^{d \times d}$ is a linear transformation matrix and ${\bf b} \in \mathbb{R}^{d}$ is a bias vector. The transformed and fixed embeddings are subsequently length-normalized and compared using cosine scoring. The parameters ${\bf W}$ and ${\bf b}$ are optimized using the same cosine embedding loss $\mathcal{L}_{\rm cosine}$ and identity regularization term $\mathcal{L}_{\rm reg}$ as in the affine cosine backend.

\section{Experimental Setup}\label{sec:experimenta-setup}
Before addressing the main task of this study, kinship verification, we first analyze speaker verification (SV). Because SV and kinship verification differ only in their trial keys, we report SV results using the curated data and protocol described in Section~\ref{sec:data}. Our goal is to determine whether \emph{strict-kin} trials constitute challenging SV non-target trials; if so, this would suggest that speaker embeddings encode kinship-related characteristics.

We then evaluate KV in both zero-shot and trainable-backend settings, as described in Section~\ref{sec:trainable_backend}. Because pretrained speaker embeddings are optimized for speaker identity, we also train backends using only \emph{strict-kin} trials, excluding same-speaker targets and encouraging the model to focus on family-related information beyond identity. Finally, inspired by gender-dependent modeling in early SV systems~\cite{reynolds2000speaker}, we also investigate the use of \emph{gender-constrained} backend training. In this approach, separate backends are trained for each gender pairing. This requires multiple gender-pair specific models and model selection at inference time, but can reduce gender-induced variability and improve kinship discrimination.



\subsection{Speaker Embeddings for Kinship Verification}
We consider three well-established pretrained speaker embedding extractors that reflect complementary architectural design paradigms for kinship verification. (1) \textbf{ECAPA-TDNN}~\cite{ecapadesplanques2020}\footnote{\url{https://github.com/TaoRuijie/ECAPA-TDNN/}} is a time delay neural network (TDNN)~\cite{waibel1989phoneme} based architecture incorporating channel-wise attention, multi-layer feature aggregation, and attentive statistical pooling, motivated by the hypothesis that speaker-discriminative information is distributed across layers and channels; it has $\sim$27.1M parameters and is trained on VoxCeleb1 and VoxCeleb2. (2) \textbf{WavLM-ECAPA}~\cite{wavlm}\footnote{\url{https://huggingface.co/microsoft/wavlm-base-sv}} integrates a self-supervised WavLM front end with an ECAPA-based embedding backend to leverage large-scale unlabeled speech representations for improved robustness and generalization; the WavLM encoder (pretrained on LibriSpeech) provides contextual features, followed by ECAPA fine-tuning on speaker verification data (e.g., VoxCeleb), with a total of $\sim$101.1M parameters. (3) 
\textbf{ReDimNet}~\cite{yakovlev24_interspeech}\footnote{\url{https://github.com/IDRnD/redimnet} } is a state-of-the-art speaker embedding model trained on VoxCeleb. It uses dimension-reshaping operations to better capture speaker-relevant characteristics and has approximately $15$M parameters.


 \subsection{Training Conditions and Inference Strategy}

The zero-shot setting involves no training procedure. For the trainable back-ends (Section~\ref{sec:trainable_backend}), we examine three alternative training conditions that differ in how the training trials are selected.
(1) \textbf{Full trials:} We use the complete set of training trials, where the target class includes both \emph{strict-kin} and \emph{speaker-target} trials, while the non-target class consists of \emph{unrelated} (kin non-target) trials. (2) \textbf{Strict kin trials:} We exclude standard speaker verification trials and train the model solely using kinship-based trials. Thus, the target class consists of \emph{strict-kin} pairs, whereas the non-target class consists of \emph{unrelated} pairs. This training condition encourages the model to focus on learning family-specific characteristics. 
(3) \textbf{Gender-constrained trials:} We further partition the strict kin training trials according to gender combinations and train separate models for each pair: female--female (FF), male--male (MM), and male--female (MF). 
Conditioning on gender encourages the model to learn kinship-related cues, rather than the more difficult task of disentangling gender and kinship effects.


During inference under the \emph{gender-constrained trials} setting, each test trial is evaluated using the corresponding gender-specific model (FF, MM, or MF). The backend then transforms the input embeddings, after which cosine similarity is computed between the transformed representations.

\subsection{Configuration of Trainable backends}

All trainable backend models are optimized for 15 epochs using the Adam optimizer with a learning rate of $10^{-4}$. For model selection, we retain the checkpoint achieving the lowest equal error rate (EER) on the validation set.

For the fully connected backend, we set the regularization coefficient to $\lambda_f = 10^{-3}$, following the configuration used in~\cite{kefalas2023kan}. For both affine cosine backends (S-AP and AS-AP), we set the regularization coefficients to $\lambda_w = 10^{-3}$ and $\lambda_b = 10^{-3}$, and use an additive margin of $m = 0.2$. All hyperparameters are selected based on validation performance by empirically testing several candidate values, with the final configuration chosen according to the lowest validation EER.

\subsection{Performance Metrics and Evaluation Conditions}


We evaluate all methods using EER for overall \textbf{KV} and strict KV (\textbf{KV$^{\ast}$}), as defined in
Section~\ref{subsec:speaker-kin-interaction}. In addition, we report category-wise EERs for each
kin relation---Sister--Sister (SS), Mother--Daughter (MD),
Brother--Brother (BB), Father--Son (FS), Brother--Sister (BS),
Mother--Son (MS), and Father--Daughter (FD)---using gender-matched
non-targets (FF, MM, MF) to control for gender effects on the score
distribution. Trial counts are shown in Fig.~\ref{fig2:Trail_Pipeline}.



\section{Experimental Results}
\subsection{Speaker Verification: Effect of Family Non-Targets}
We first report the zero-shot SV performance of ECAPA-TDNN, WavLM-ECAPA, and ReDimNet embeddings in Table~\ref{tab:SV_ZS}. 
The evaluation considers three non-target conditions: \emph{strict-kin}, \emph{unrelated}, and their pooled combination. ReDimNet consistently outperforms the other speaker embedding extractors, obtaining 
pooled EER of $5.41\%$. 

\begin{figure}[t]
\centering
\includegraphics[height= 100pt,width=220pt]{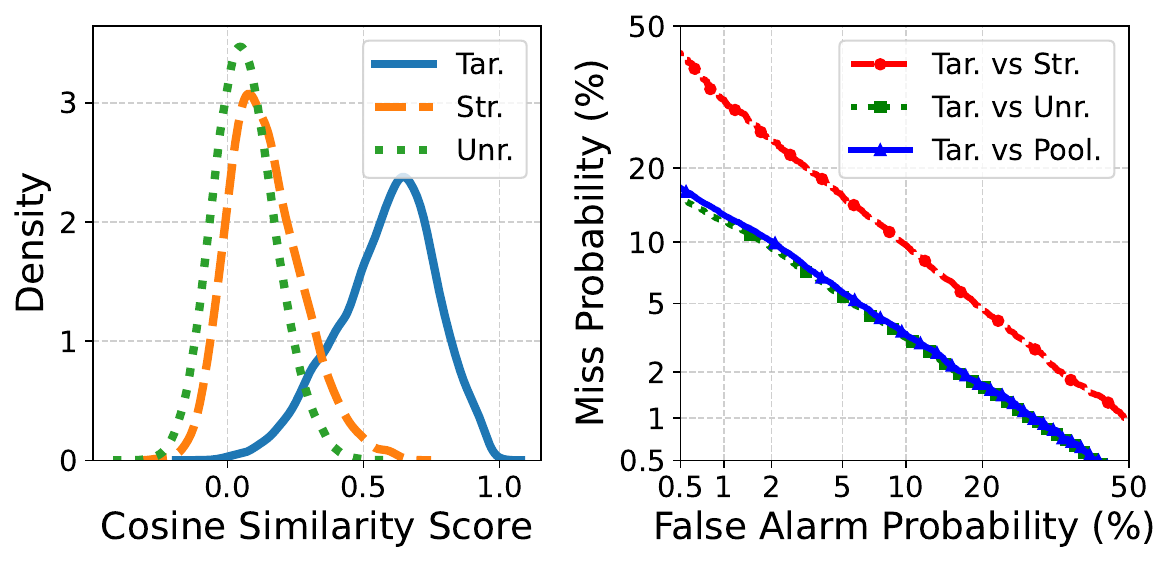}
  \caption{Score distributions and DET plot of RedimNet SV system with speaker-target (Tar.), strict-kin (Str.), and unrelated (Unr.) trials.}
\label{fig:SD_SV}
\end{figure}

As expected, SV performance degrades considerably when \emph{strict-kin} trials are used as non-targets, compared with the \emph{unrelated} condition. For example, ReDimNet yields an EER of $9.83\%$ on \emph{strict-kin} trials, compared with $5.26\%$ on \emph{unrelated} trials. Beyond confirming that family members provide a more challenging testbench for SV, this observation suggests that speaker embeddings capture certain kinship-related characteristics. Consistently, both the score distributions and detection estimation tradeoff (DET) profiles displayed in Fig.~\ref{fig:SD_SV} indicate that \emph{strict-kin} non-targets overlap more substantially with target trials than \emph{unrelated} non-targets do. 
At any fixed false-alarm rate, \emph{strict-kin} trials consistently produce higher miss rates than \emph{unrelated} trials.

\begin{figure}[t]
\centering
\includegraphics[height= 140pt,width=240pt]{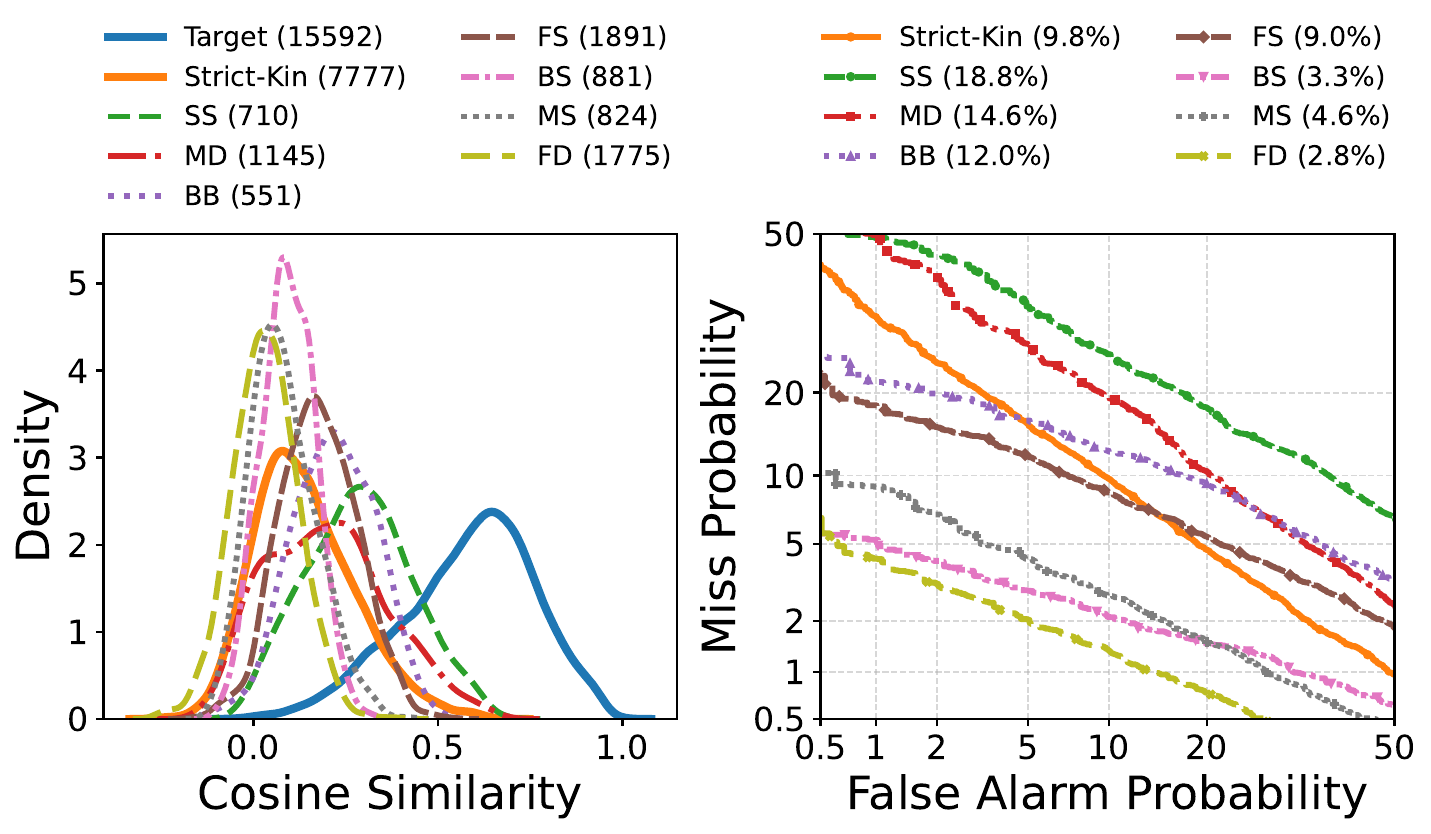}
  \caption{Score distributions (SD) and DET plots of the RedimNet SV system for speaker-target (Target) and various subcategories of strict-kin trials. SD plots, parentheses indicate the number of trials, DET plots, they indicate EER.}
\label{fig:SD_KDE_Sub_SV}
\end{figure}

We further observe that the pooled-setting EER ($5.41\%$ with ReDimNet) remains closer to the performance obtained with \emph{unrelated} non-targets. This behavior is primarily explained by the composition of the pooled non-target set, in which approximately $96.6\%$ of the trials are \emph{unrelated} pairs and only about $3.4\%$ are \emph{strict-kin} pairs. As a more fine-grained analysis of the latter subset, Fig.~\ref{fig:SD_KDE_Sub_SV} breaks down the performance by the kin relation. It is evident that this factor strongly influences the results. As expected, same-gender pairs such as SS (sister--sister) and BB (brother--brother) are more confusable than mixed-gender pairs such as BS (brother--sister). Overall, the strict-kin condition lies closer to the target distribution than the unrelated nontarget conditions, resulting in greater score overlap and higher EERs. This suggests that familial similarity, especially within same-gender kin pairs, increases the difficulty of SV.


\begin{table}[t]
\centering
\small
\caption{SV Performance (EER) across embeddings for strict kin, unrelated, and pooled Non-target trials. Lower is better.}
\begin{tabular}{l|ccc}
\toprule
\textbf{Trials} & \textbf{Strict Kin } & \textbf{Unrelated } & \textbf{Pooled} \\
$\%$ trials  & $3.4\%$ & $96.6\%$ & $100\%$ \\
\midrule
ECAPA         & 13.02 & 7.35 & 7.58 \\
WavLM-ECAPA   & 21.36 & 17.45 & 17.57 \\
RedimNet      & \textbf{9.83} & \textbf{5.26} & \textbf{5.41} \\
\bottomrule
\end{tabular}
\label{tab:SV_ZS}
\end{table}

\subsection{Zero-shot kinship verification}

Moving to kinship verification, the zero-shot results 
are shown in Table~\ref{tab:ZS_KV}. Consistent with the SV results, ReDimNet achieves the best overall KV performance with an EER of $20.8\%$, followed by ECAPA-TDNN ($21.0\%$) and WavLM-ECAPA ($28.2\%$). These EERs are far higher than the corresponding SV EERs. This can be expected, since that embedding extractor was trained for SV rather than KV.

\begin{table}[t]
\centering
\small
\caption{KV Performance (EER) comparison with zero-shot setting.} 
\label{tab:ZS_KV}
\small
\begin{tabular}{l|
>{\columncolor{cyan!20}}c >{\columncolor{cyan!20}}c
>{\columncolor{yellow!20}}c >{\columncolor{yellow!20}}c
>{\columncolor{green!40}}c >{\columncolor{green!40}}c >{\columncolor{green!40}}c
c c}

\toprule
\textbf{Emb.} &
\multicolumn{2}{c}{\textbf{FF}} &
\multicolumn{2}{c}{\textbf{MM}} &
\multicolumn{3}{c}{\textbf{MF}} &
\textbf{KV*} & \textbf{KV} \\
\cmidrule(lr){2-3} \cmidrule(lr){4-5} \cmidrule(lr){6-8}

& SS & MD
& BB & FS
& BS & MS & FD
&& \\
\midrule

ECAPA
& 22.8 & 33.7
& 30.5 & 29.3
& 33.4 & 42.5 & 47.0
& 38.4 & 21.0 \\

WavLM-E
& 27.9 & 38.9
& 34.1 & 35.2
& 52.4 & 29.2 & 53.2
& 49.3 & 28.2 \\

RedimNet
& 22.1 & 34.0
& 25.6 & 31.1
& 32.8 & 38.7 & 48.0
& 39.7 & \textbf{20.8} \\
\bottomrule
\end{tabular}
\end{table}

We further compare the standard KV setting with the strict KV* setting. In KV*, where speaker-target trials are excluded from the target set, EERs increase substantially: from $20.8\%$ to $39.7\%$ for ReDimNet and from $21.0\%$ to $38.4\%$ for ECAPA-TDNN. This indicates that including speaker-target trials in the standard KV protocol makes the task easier, since pretrained speaker embeddings are inherently more discriminative for speaker identity than for kinship. 

Finally, ReDimNet outperforms ECAPA-TDNN in the standard KV setting, whereas ECAPA-TDNN ($38.4\%$ EER) marginally outperforms ReDimNet ($39.7\%$ EER) in the stricter KV* setting.  Similar to the SV results, WavLM-ECAPA remains inferior to the rest in both KV and KV*. Since ReDimNet achieves better performance across most kinship categories, we select it as the primary embedding extractor for subsequent experiments with trainable backends. For completeness, we also report results using ECAPA-TDNN with the best-performing trainable backend configuration.

Using ReDimNet embeddings, we further analyze performance across the kin relation. The SS pairs achieve the lowest EER among all relations. In general, performance degrades with both increasing the genological level and cross-gender pairings. In particular, for a given gender configuration (FF, MM, or MF), increasing the genological level leads to a consistent deterioration in performance. Conversely, for a fixed genological level, cross-gender relations yield higher EERs compared to same-gender relations (e.g., SS and BB outperform BS; similarly, MD and FS outperform MS and FD). These trends indicate that both gender mismatch and the type of familial relationship jointly influence KV performance.













\begin{table}[t]
\centering
\small
\caption{KV Performance Using Trainable Backends with \textbf{Full Trials}. ZS: zero-shot (RedimNet), FCN: Fully Connected Network.}
\label{tab:KV_TB_FT}
\begin{tabular}{c|
>{\columncolor{cyan!20}}c >{\columncolor{cyan!20}}c
>{\columncolor{yellow!20}}c >{\columncolor{yellow!20}}c
>{\columncolor{green!40}}c >{\columncolor{green!40}}c >{\columncolor{green!40}}c
c c}

\toprule

\textbf{Model} &
\multicolumn{2}{c}{\textbf{FF}} &
\multicolumn{2}{c}{\textbf{MM}} &
\multicolumn{3}{c}{\textbf{MF}} &
\textbf{KV*} &
\textbf{KV} \\

\cmidrule(lr){2-3} \cmidrule(lr){4-5} \cmidrule(lr){6-8}

&
SS & MD &
BB & FS &
BS & MS & FD &
& \\

\midrule


\textbf{FCN} &
26.2 & 51.7 &
32.5 & 41.8 &
42.9 & 39.1 & 47.7 &
44.2 & 29.9 \\ 

\textbf{S-AP} &
20.0 & 35.6 &
28.7 & 32.6 &
34.6 & 32.0 & 51.9 &
\textbf{37.2} & \textbf{20.3} \\ 


\bottomrule
\end{tabular}
\end{table}

\subsection{Kinship verification with trainable backends}
Turning to KV with trainable backends, we evaluate the fully connected (FCN) Siamese backend and the proposed symmetric affine projection (S-AP) backend under the \emph{Full Trials} training condition. The results shown in Table~\ref{tab:KV_TB_FT} 
indicate that FCN 
performs \emph{worse} than the zero-shot baseline for both KV and KV*:
the EER increases from $20.8\%$ to $29.9\%$ for KV and from $39.7\%$ to $44.2\%$ for KV*. 
We attribute this decline to overfitting due to the limited amount of backend training data, which likely hinders generalization to the evaluation data.


In contrast, the proposed S-AP backend consistently improves performance over both the zero-shot baseline and the FCN backend. Compared to the FCN backend, S-AP uses substantially fewer trainable parameters while preserving the cosine similarity structure through identity regularization. It achieves EERs of $20.3\%$ for KV and $37.2\%$ for KV$^\ast$, improving with a statistically significant margin\footnote{Significance was evaluated using the parametric HTER-based $z$-test with $95\%$ confidence interval following~\cite{bengio2004statistical}. As the reported metric is EER, the test was applied at the EER operating point by approx. $\mathrm{FAR}\approx\mathrm{FRR}\approx\mathrm{EER}$.} over the corresponding zero-shot baselines of $20.8\%$ ($p=0.01$) and $39.7\%$ ($p=2.89\times10^{-10}$), respectively. These results indicate that the proposed trainable symmetric affine projection, together with identity regularization, is more effective than the FCN-based backends for KV under data-constrained conditions.

\subsection{Affine projection backend with strict-kin  trials training}

\label{sec:strict_kin_results}

\begin{table}[t]
\centering
\small
\caption{KV Performance (EER) with \textbf{Strict Kin} trials.}
\label{tab:KO-T}
\begin{tabular}{c|c|
>{\columncolor{cyan!20}}c >{\columncolor{cyan!20}}c
>{\columncolor{yellow!20}}c >{\columncolor{yellow!20}}c
>{\columncolor{green!40}}c >{\columncolor{green!40}}c >{\columncolor{green!40}}c
c c}

\toprule

\textbf{Model} &
\textbf{Bias} &
\multicolumn{2}{c}{\textbf{FF}} &
\multicolumn{2}{c}{\textbf{MM}} &
\multicolumn{3}{c}{\textbf{MF}} &
\textbf{KV*} &
\textbf{KV} \\

\cmidrule(lr){3-4} \cmidrule(lr){5-6} \cmidrule(lr){7-9}

& &
SS & MD &
BB & FS &
BS & MS & FD &
& \\

\midrule


\textbf{S-AP} &  N &
23.5 & 34.5 &
30.1 & 32.3 &
36.1 & 32.6 & 47.3 &
36.6 & 20.2 \\ \midrule

\multirow{2}{*}{\textbf{AS-AP}} & N &
18.6 & 34.1 &
28.2 & 34.9 &
30.5 & 30.8 & 40.5 &
32.8 & 19.3 \\ \cmidrule{2-11}
 & Y &
19.1 & 33.9 &
28.4 & 34.5 &
31.0 & 30.0 & 40.7 &
\textbf{32.6} & \textbf{19.3} \\ 

\bottomrule
\end{tabular}
\end{table}

\begin{table}[t]
\centering
\small
\caption{KV Performance (EER) using asymmetric affine projection (AS-AP) with Strict Kin Trials with various ordering.}
\label{tab:KO-T_AS_ordering}
\begin{tabular}{c|
>{\columncolor{cyan!20}}c >{\columncolor{cyan!20}}c
>{\columncolor{yellow!20}}c >{\columncolor{yellow!20}}c
>{\columncolor{green!40}}c >{\columncolor{green!40}}c >{\columncolor{green!40}}c
c c}

\toprule

\multirow{2}{*}{\textbf{Emb. Ord.}} &

\multicolumn{2}{c}{\textbf{FF}} &
\multicolumn{2}{c}{\textbf{MM}} &
\multicolumn{3}{c}{\textbf{MF}} &
\textbf{KV*} &
\textbf{KV} \\

\cmidrule(lr){2-3} \cmidrule(lr){4-5} \cmidrule(lr){6-8}

&
SS & MD &
BB & FS &
BS & MS & FD &
& \\

\midrule
\textbf{-} & 
18.7 & 30.6 &
31.3 & 33.1 &
33.7 & 28.4 & 43.5 &
33.7 & 19.5 \\ 

\textbf{age} & 
18.6 & 34.1 &
28.2 & 34.9 &
30.5 & 30.8 & 40.5 &
\textbf{32.8} & \textbf{19.3} \\ 

\textbf{gen} &
18.3 & 34.0 &
29.5 & 34.6 &
33.4 & 42.1 & 41.6 &
35.2 & 20.2 \\ 

\textbf{age \& gen} &
19.8 & 37.8 &
25.8 & 34.0 &
34.1 & 43.3 & 39.9 &
34.9 & 20.2 \\ 

\bottomrule
\end{tabular}
\end{table}

Next, we employ affine projection in both symmetric and asymmetric formulations (Section~\ref{sec:kinship-verification}) trained 
only with \textbf{strict-kin} trials. 
The results in Table~\ref{tab:KO-T} show improvements for both KV and KV*, with EERs of $20.2\%$ and $36.6\%$, respectively, compared with $20.3\%$ and $37.2\%$ under the all-trials training condition. This indicates that restricting backend training to strict-kin trials helps the model capture kinship-relevant characteristics more effectively. 

We further investigate asymmetric affine projection (AS-AP) using different embedding ordering strategies as explained in Section~\ref{sec:trainable_backend}. The results in Table~\ref{tab:KO-T_AS_ordering} indicate that all asymmetric projection strategies outperform the symmetric projection baseline under both KV and KV* conditions. This trend is consistent across all ordering schemes. Age-based ordering achieves the best results, statistically reducing the EER to $19.3\%$ for KV and $32.8\%$ for KV*, compared with $20.2\%$ ($p=3.2\times 10^{-6}$) and $36.6\%$ ($p\approx 0$), respectively, for symmetric projection (S-AP). These results demonstrate the effectiveness of asymmetric projection.

Age-based ordering consistently outperforms the unordered condition, whereas gender-based ordering and sequential age--gender ordering perform worse. This is especially clear for cross-gender pairs (BS, MS, FD), where projecting the older speaker's embedding into the younger speaker's space outperforms gender-based ordering. The results indicate that age-related variability outweighs gender mismatch in asymmetric projection learning. Based on these findings, we adopt age-based ordering in all subsequent AS-AP experiments. The corresponding results are included in Table~\ref{tab:KO-T} for comparison.

Up to this point, the affine projection backends only learn scaling weights without any bias term. We now extend the AS-AP backend by incorporating an additive bias term together with age-based ordering.  Adding a bias term marginally reduces KV* EER from $32.8\%$ to $32.6\%$ (not significant, $p=0.6$), suggesting that the added flexibility provides only a marginal benefit.

\subsection{Affine projection backend with gender constraint  trials}

\begin{table}[t]
\centering
\small
\caption{KV Performance (EER) with \textbf{gender constrained} trials}
\label{tab:GC-T_AP}
\begin{tabular}{c|c|
>{\columncolor{cyan!20}}c >{\columncolor{cyan!20}}c
>{\columncolor{yellow!20}}c >{\columncolor{yellow!20}}c
>{\columncolor{green!40}}c >{\columncolor{green!40}}c >{\columncolor{green!40}}c
c c}

\toprule

\textbf{Model} &
\textbf{Bias} &
\multicolumn{2}{c}{\textbf{FF}} &
\multicolumn{2}{c}{\textbf{MM}} &
\multicolumn{3}{c}{\textbf{MF}} &
\textbf{KV*} &
\textbf{KV} \\

\cmidrule(lr){3-4} \cmidrule(lr){5-6} \cmidrule(lr){7-9}

& &
SS & MD &
BB & FS &
BS & MS & FD &
& \\

\midrule


\textbf{S-AP} &  N &
22.0 & 35.7 &
26.3 & 31.9 &
45.2 & 31.3 & 43.9 &
34.9 & 19.5 \\ \midrule

\multirow{2}{*}{\textbf{AS-AP}} & N &
18.7 & 34.9 &
26.7 & 33.7 &
36.4 & 33.6 & 42.8 &
32.3 & 18.8 \\  \cmidrule{2-11}

 & Y &
17.6 & 35.7 &
26.3 & 33.8 &
36.3 & 33.6 & 42.5 &
\textbf{32.0} & \textbf{18.6} \\ 
\bottomrule
\end{tabular}
\end{table}


\begin{figure}[t]
\centering
\includegraphics[height= 110pt,width=150pt]{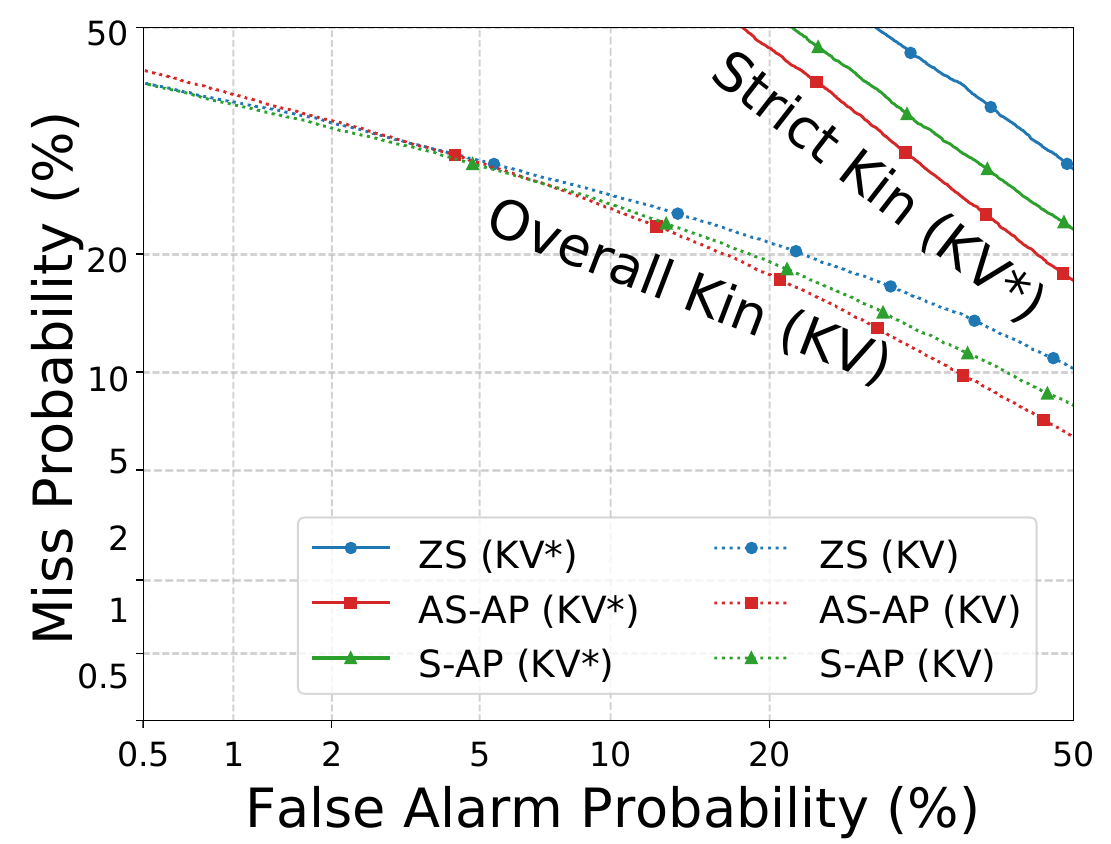}
  \caption{DET curve: comparison with zero-shot and trainable backends. }\label{fig:DET_KV}
\end{figure}

We now turn our attention to KV under the \textbf{gender-constrained} training setup detailed in Section~\ref{sec:experimenta-setup}. 
The results for both the symmetric and asymmetric trainable projection methods (Table~\ref{tab:GC-T_AP}) indicate consistent performance improvement from gender-agnostic to gender-dependent models, regardless of system configuration. The best system, AS-AP with bias, achieves EERs of $32.0\%$ on KV$^\ast$ and $18.6\%$ on KV, suggesting improvement over the gender-agnostic strict-kin configuration in Table~\ref{tab:KO-T} ($32.6\%$ and $19.3\%$, respectively). However, only the KV improvement is statistically significant ($p=2.0\times10^{-4}$); the KV$^\ast$ gain is not ($p=0.11$). Therefore, the advantage of gender-constrained training remains limited.  

Fig.~\ref{fig:DET_KV} further illustrates the corresponding DET profiles. For KV$^\ast$, the miss rate at any fixed FAR decreases from the zero-shot baseline to symmetric projection (S-AP), and decreases further with the proposed asymmetric projection (AS-AP) using age-based ordering, indicating improved detection performance across operating points. 
For overall KV, a similar improvement is visible mainly beyond about $5\%$ FAR, whereas the curves overlap in the lower-FAR region. 
Moreover, the KV$^\ast$ curves remain consistently above the corresponding KV curves. This shows that removing speaker-target trials makes the verification problem more challenging. 



Finally, replacing RedimNet with ECAPA-TDNN under the best backend
(Table~\ref{tab:KV_emb}) shows the same trend as in SV: RedimNet outperforms
ECAPA-TDNN ($p \approx 0$ for both KV and KV$^{\ast}$), suggesting it carries
richer kin
 related information that the asymmetric projection can
exploit effectively.

\begin{table}[t]
\centering
\small
\caption{KV: various embeddings using gender constrained training.}
\label{tab:KV_emb}
\begin{tabular}{c|c|
>{\columncolor{cyan!20}}c >{\columncolor{cyan!20}}c
>{\columncolor{yellow!20}}c >{\columncolor{yellow!20}}c
>{\columncolor{green!40}}c >{\columncolor{green!40}}c >{\columncolor{green!40}}c
c c}

\toprule

\textbf{Emb.} &
\textbf{Bias} &
\multicolumn{2}{c}{\textbf{FF}} &
\multicolumn{2}{c}{\textbf{MM}} &
\multicolumn{3}{c}{\textbf{MF}} &
\textbf{KV*} &
\textbf{KV} \\

\cmidrule(lr){3-4} \cmidrule(lr){5-6} \cmidrule(lr){7-9}

& &
SS & MD &
BB & FS &
BS & MS & FD &
& \\

\midrule


\textbf{RedimNet} &  Y &
17.6 & 35.7 &
26.3 & 33.8 &
36.3 & 33.6 & 42.5 &
\textbf{32.0} & \textbf{18.6} \\ \midrule

\textbf{ECAPA} & Y &
20.8 & 34.1 &
30.1 & 34.5 &
34.8 & 40.5 & 46.5 &
35.3 & 20.7 \\ 

\bottomrule
\end{tabular}
\end{table}

\section{Discussion}

Before concluding, we revisit the role of age and type of kinship relation to KV performance.

\subsection{The Impact of Age Difference}

Our results indicate that age remains an important source of performance variation. Additional analysis using the best-performing gender-constrained affine projection system (Table~\ref{tab:age_diff}) shows that stricter age-difference constraints improve performance. When both target and non-target trials are limited to a maximum age difference of $\leq\!5$ or $\leq\!2$ years, EERs decrease accordingly: KV* EER drops from $32.0\%$ (no constraints) to $29.3\%$ and further to $27.9\%$. This confirms that larger age differences between related speakers make KV more challenging.

Applying age-difference constraints to both target and non-target trials yields a controlled test of kinship-specific discrimination. Constraining only the target trials, however, might better reflect practical deployment; while the age gap between enrollment and target-side kin samples can be controlled when longitudinal family enrollment speech is available, the age distribution of non-target trials remains uncontrolled. We therefore further evaluate strict KV* under target-only age-difference constraints, using the zero-shot, symmetric, and asymmetric affine projection backends trained with gender-constrained partitions.



\begin{figure}[t]
\centering
\includegraphics[height= 130pt,width=190pt]{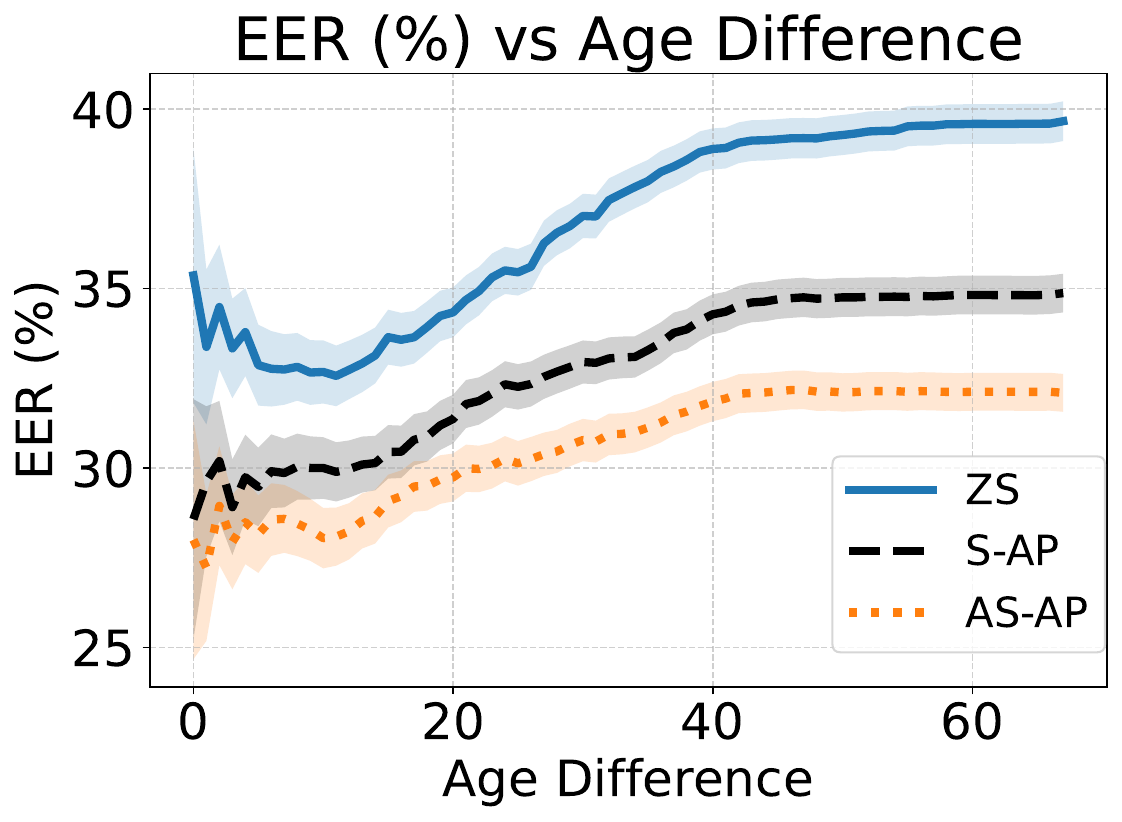}
   \caption{EER of KV$^{*}$ systems with constrained ($\leq$) target-trial age difference; shaded regions denote the $\pm\,95\%$ confidence interval.}\label{fig:eer_age_KV}
\end{figure}

As Fig.~\ref{fig:eer_age_KV} shows, EER increases with age difference, apart from small fluctuations in the low age-difference region, likely due to the small sample size. Across all age differences, the asymmetric affine projection consistently outperforms both the symmetric projection and the zero-shot methods. 
Notably, the performance of all systems stabilizes for age differences above $40$ years. 
Overall, both the absolute EER and its variation across the age-difference axis are smaller for the asymmetric affine projection than for the symmetric affine projection and the zero-shot method. This indicates that the proposed method reduces, but does not fully eliminate, the impact of age difference.

\begin{table}[t]
\centering
\small
\caption{EER (\%): age-difference (Age Diff.) constraints applied
to both target and non-target trials. ``$-$:'' unconstrained.}
\label{tab:my-table}

\begin{tabular}{c|
>{\columncolor{cyan!20}}c >{\columncolor{cyan!20}}c
>{\columncolor{yellow!20}}c >{\columncolor{yellow!20}}c
>{\columncolor{green!40}}c >{\columncolor{green!40}}c >{\columncolor{green!40}}c
c c}

\toprule

\textbf{Age Diff} &
\multicolumn{2}{c}{\textbf{FF}} &
\multicolumn{2}{c}{\textbf{MM}} &
\multicolumn{3}{c}{\textbf{MF}} &
\textbf{KV*} &
\textbf{KV} \\

\cmidrule(lr){2-3} \cmidrule(lr){4-5} \cmidrule(lr){6-8}

&
SS & MD &
BB & FS &
BS & MS & FD &
& \\
\midrule
\textbf{-} &
\textbf{17.6} & 35.7 &
\textbf{26.3} & 33.8 &
36.3 & \textbf{33.6} & 42.5 &
\textbf{32.0} & 18.6 \\ \midrule

\textbf{$\leq$ 5} &
\textbf{16.4} & 37.2 &
\textbf{27.1} & 34.1 &
33.8 & \textbf{29.2} & 38.8 &
\textbf{29.3} & 12.5 \\ 

\textbf{$\leq$ 2} &
\textbf{20.2} & 35.8 &
\textbf{25.0} & 29.5 &
33.6 & \textbf{29.5} & 36.8 &
\textbf{27.9} & 9.7 \\ 

\bottomrule
\end{tabular}
\label{tab:age_diff}
\end{table}

\subsection{Relative Difficulty of Kinship Subcategories}


As a final analysis, we examine the relative difficulty of kinship subcategories in speech-based KV. As shown in Table~\ref{tab:age_diff}, the proposed system exhibits a consistent relation-wise ranking across all three gender-pair conditions. For same-gender FF pairs, SS trials are easier than MD trials. Similarly, for MM pairs, BB trials are easier than FS trials. This suggests that, under gender-matched conditions, same-generation sibling relations may be easier than cross-generation parent--child relations. 
However, the mixed-gender relations show a less uniform pattern. In our results, MS trials are easier than BS and FD trials. 


It is also instructive to compare our findings to the earlier scarce literature. Because prior studies use different datasets, protocols, and metrics, we compare only relation-wise difficulty \emph{rankings} rather than absolute performance. Even relative rankings are protocol-sensitive, so our comparison should be read in light of the family-disjoint split and the non-parametric standardization of age difference and gender composition used in our protocol. The resulting ordering is broadly stable under additional age-difference constraints, but only partially agrees with earlier KAN-AV studies~\cite{kefalas2023kan,sun2024audio}. FF relations show similar rankings across studies, whereas MM relations differ, with prior work ranking FS easier than BB, unlike our results. For mixed-gender relations,~\cite{kefalas2023kan} is the same as our ordering, while~\cite{sun2024audio} ranks BS easier than MS and FD. These differences might reflect dataset partitioning, protocols, and residual confounding. Relation-wise EERs should therefore be interpreted as diagnostic indicators of difficulty, not as definitive measures of the intrinsic acoustic strength of each kin relation.



Taken together, these findings highlight an important difference between SV and KV. In SV, demographic variability such as age and gender mismatch can often be reduced by trial design, for example, by constructing matched-gender or age-difference controlled non-target trials. In KV, however, age and gender differences are inherently tied to many target relations and therefore cannot simply be removed without changing the nature of the verification task. For example, FS and MD trials naturally involve cross-generation age differences, while MS, FD, and BS trials necessarily involve cross-gender comparisons. Thus, while age and gender are not kinship evidence by themselves, they \emph{interact} strongly with the relational structure that defines the KV problem. The proposed asymmetric affine projection provides a first step toward handling this interaction by allowing one side of the embedding pair to be transformed according to an ordered relation. 
Future work should study kinship verification together with the structured effects of age and gender, rather than treating them only as nuisance factors to be suppressed. A deeper understanding of which voice cues are genuinely kinship-relevant requires more controlled acoustic analysis using phonetically balanced, high-quality, and noise-free family speech data.

\section{Conclusion}
\label{con}

We formulated voice-based kinship verification as an open-world pairwise detection task linked to speaker verification, with strict-kin trials isolating familial similarity beyond speaker identity. Using a curated, family-disjoint KAN-AV protocol with matched gender and age-difference distributions, we showed that neural speaker embeddings encode measurable kinship cues. The overall EERs, however, are about an order of magnitude higher than in speaker verification studies. Our results indicate that age difference remains a major source of variation: stricter age constraints improve performance. The proposed asymmetric projection reduces, but does not entirely disentangle, age-related confounding. Despite these limitations, our work provides a foundational basis for further study into kinship analysis using voice. Future work should move toward larger controlled family speech corpora collection and improved disentanglement of kinship, age and gender cues, in addition to analyzing acoustic correlates of kinship.

\bibliographystyle{IEEEtran}
\bibliography{sampbib}

\vspace{11pt}


\vfill

\end{document}